# Prediction of a wide variety of linear complexions in face centered cubic alloys

Vladyslav Turlo [a], Timothy J. Rupert [a,b,*]

[a] Department of Mechanical and Aerospace Engineering, University of California, Irvine, CA 92697, USA
[b] Department of Materials Science and Engineering, University of California, Irvine, CA 92697, USA
* Email: trupert@uci.edu

Linear complexions are defect states that have been recently discovered along dislocations in body centered cubic Fe-based alloys. In this work, we use atomistic simulations to extend this concept and explore segregation-driven structural transitions at dislocations in face centered cubic alloys. We identify a variety of stable, nanoscale-size structural and chemical states, which are confined near dislocations and can be classified as linear complexions. Depending on the alloy system and thermodynamic conditions, such new states can preserve, partially modify, or completely replace the original defects they were born at. By considering different temperatures and compositions, we construct *linear complexion diagrams* that are similar to bulk phase diagrams, defining the important conditions for complexion formation while also specifying an expected complexion size and type. Several notable new complexion types were predicted here: (1) nanoparticle arrays comprised of $L1_2$ phases in Ni-Fe, Ni-Al, and Al-Zr, (2) replacement of stacking faults with layered complexions comprised of (111) planes from the $Cu_5Zr$ intermetallic phase in Cu-Zr, (3) platelet arrays comprised of two-dimensional Guinier-Preston zones in Al-Cu, and finally (4) coexistence of multiple linear complexions containing both Guinier-Preston zones and $L1_2$ phases in ternary Al-Cu-Zr. All of these new complexion states are expected to alter material properties and affect the stability of the dislocations themselves, offering a unique opportunity for future materials design.

## 1. Introduction

*Complexions* are thermodynamically-stable chemical and structural states confined to crystalline defects [1-4]. To date, the vast majority of research on complexions has been mainly focused on *planar complexions*, or two-dimensional defect states at interfaces such as grain boundaries, interphases, or surfaces [3-14]. With varying thermodynamic parameters such as temperature and composition, interfacial regions may abruptly change their structure and composition, switching among complexion types in a manner that is similar to a bulk phase transition [9-18]. The differentiating factor is that a complexion transition is dependent on the defect to locally alter the crystal structure, stress, composition, or another thermodynamic variable in a way that stabilizes the nanoscale phase in the local environment. Such complexion transitions may significantly affect the mechanical [19-22] and atomic transport [9] properties as well as thermal stability [15, 22-24] of polycrystalline materials. For example, a transition to disordered nanolayer complexions, also be called amorphous intergranular films, improves both the strength and ductility of nanocrystalline Cu-Zr [20]. Disordered complexions can accommodate an unexpected region of high-temperature nanocrystalline stability against grain growth [24]. As another example, the transition from a monolayer to bilayer complexions is able to explain an abrupt change in activation energy for grain boundary diffusion in the Cu-Ag system [9].

In contrast to the more widely investigated planar complexions, *linear complexions* (one-dimensional defect states confined along a dislocation line) have only been recently discovered. The term linear complexion was first proposed in the recent experimental work of Kuzmina et al. [1] to describe arrays of nanoscale precipitates of the austenite phase that were formed along the dislocation lines in body centered cubic Fe-9 at.% Mn. Kwiatkowski da Silva et al. [25] subsequently performed a thermodynamic analysis of the Fe-Mn system, suggesting a novel phase transformation path of spinodal decomposition in a defect segregation zone followed by austenite phase nucleation in the regions with elevated composition. Such regions limit the growth of the precipitates, explaining the formation of the nanoscale precipitate arrays along dislocations. Despite such progress, a systematic experimental investigation of alloys in relation to temperature and composition is very difficult, while thermodynamic models require many important parameters to be defined ahead of time, slowing the identification of new types of linear complexions.

On the other hand, atomistic simulations have proven to be a powerful tool for investigating complexion formation phenomenon, as they capture all of the chemical and structural information associated with nanoscale phase transformations. For example, atomistic models have provided great insight into grain boundary complexion transformations such as a "split-kite"-to-"filled-kite" transitions at tilt grain boundaries in Cu-Ag [10], ordered-to-disordered complexion transitions and the formation of nanoscale amorphous intergranular films in Cu-Zr [16], and a symmetric-to-asymmetric structural transition at Mo-Ni tilt grain boundaries [17]. Yang et al. [17] were even able to construct a grain boundary complexion (phase) diagram by finding the transition line and critical inflection point, which would require an enormous amount of time and effort if tackled experimentally. Moreover, atomistic simulations have recently been used to study linear complexion formation in the body centered cubic Fe-Ni system by Turlo and Rupert [26]. These authors chose the Fe-Ni system because its bulk phase diagram was similar to that of Fe-Mn system, where linear complexions were first discovered, and due to the availability of a reliable Fe-Ni interatomic potential. Investigating the solute segregation and second-phase precipitation for a wide range of compositions and temperatures, Turlo and Rupert constructed a linear complexion phase diagram to describe the conditions needed to enable a complexion transition.

Nucleation of nanoscale precipitate arrays, comprised of metastable B2-FeNi and stable $L1_0$-FeNi, were uncovered. In later work, these same authors [27] demonstrated that, due to the low interfacial energy with the matrix phase, the metastable B2-FeNi phase is more thermodynamically-favorable at the nanoscale. The metastable phase is formed first, occupying the dislocation segregation zone, and then the stable phase precipitates nucleate inside of the metastable phase, resulting in the linear complexion structure with dual phase coexistence that was observed in atomistic simulations [26, 27].

A common aspect of the existing literature on linear complexions is that it is focused on body centered cubic metals [1, 25, 26, 28]. While linear complexions in face centered cubic metals have not yet been reported, they should be different from those in body centered cubic metals. Dislocations in face centered cubic metals are typically made of two Shockley partial dislocations with a stacking fault between them, providing a fundamentally different template for segregation and linear complexion formation. Since all dislocation dissociate into Shockley partials and all Shockley partials have an edge component, all dislocations will have a hydrostatic stress to promote local solute segregation. Moreover, solute absorption by stacking faults or so-called Suzuki segregation [29] may also be active. Suzuki segregation is not driven by a local stress field but instead results from the different local crystal structure and is strongly dependent on the degree of supersaturation [30], affecting the dissociation width of extended dislocations in face centered cubic alloys [31]. Using modern high-resolution electron microscopy and atom-probe tomography, recent experimental works report atomic-scale evidence for more complex segregation patterns and dislocation behavior [32-34]. For example, Kontis et al. [32] investigated Ni-based superalloys with Al, Ti, Co, and Cr addition, finding Co and Cr segregation to dislocations. This segregation provided a path for a pipe diffusion while attached to the $\gamma/\gamma'$ interface, leading to complete dissolution of the $\gamma'$ precipitates. Another study of Ni-based superalloys performed by Smith et al. [34] shows evidence of solute segregation to superlattice intrinsic stacking faults followed by a nanoscale $\gamma'$-to-$D0_{19}$ phase transformation. Investigating CoNi-based superalloys, Makineni et al. [33] reported the same nanoscale $\gamma'$-to-$D0_{19}$ transformation, while provided better chemical mapping using three-dimensional atom-probe tomography. These authors demonstrated that the stacking fault is enriched with W and Co, while the corresponding leading partial dislocation is enriched with Cr and Co and creates Cottrell atmosphere in the adjoining planes ahead that are enriched with Al [33]. Thus, by demonstrating different segregation patterns, local compositions and structures, stacking faults and also partial dislocations should be considered in some cases as independent candidates for linear complexion formation.

In this study, we use atomistic modeling to investigate the local structure and chemistry near dislocations under a wide range of compositions and temperatures, predict a variety of linear complexions, and construct the corresponding linear complexion diagrams for multiple face centered cubic alloys. First, we study the Ni-Fe system, where we find linear complexions in the form of nanoparticle arrays made of an intermetallic $L1_2$-FeNi$_3$ phase. Similar complexions were also found in the Ni-Al and Al-Zr systems. Next, we observe Zr segregation to partial dislocations and stacking faults in the Cu-Zr system that leads to the formation of both ordered and disordered complexion states. We also uncover the formation of platelet precipitates that resemble Guinier-Preston (GP) zones in Al-Cu that experience directional growth starting at partial dislocations following by faceting of the dislocation line. Finally, we demonstrate the formation of linear complexion compounds comprised of both nanoscale precipitate arrays ($L1_2$-Al$_3$Zr phase) and GP zone platelets in the ternary Al-Zr-Cu system. As a whole, our modeling results indicate that a vast material design space is available for linear complexion engineering.

## 2. Methods

Atomistic simulations were performed with the Large-scale Atomic/Molecular Massively Parallel Simulator (LAMMPS) software [35] using many-body interatomic potentials for binary systems Ni-Fe [36], Cu-Zr [37], Ni-Al [38], Al-Zr [39], Al-Cu [39], and for ternary Al-Cu-Zr [39]. While the Ni-Fe potential has already been used to investigate linear complexion formation in body centered cubic alloys [26] (i.e., in the Fe-rich side of the phase diagram), the potentials for the Cu-Zr, Al-Zr, and Al-Cu systems have all been used to capture planar complexion transitions at grain boundaries in face centered cubic alloys in prior studies [16, 40, 41]. Since the structure and chemistry of the nanoscale complexions were of primary interest, interatomic potentials were chosen that accurately recreate the major features of the experimental phase diagrams such as stable second phases and solubility limits. Equilibrium states were obtained with hybrid Monte Carlo/molecular dynamics simulations using a parallel algorithm proposed by Sadigh et al. [42]. One Monte Carlo step was performed for every 100 molecular dynamics time steps. A variance-constrained semi-grand canonical ensemble was used during the Monte Carlo portion to target a specific composition, while structural relaxation during the molecular dynamics equilibration procedure allowed for phase or complexion transformation. Each initial system was first heated from 0 K to a target temperature over 0.1 ns and then relaxed at this temperature for another 0.1 ns. A Nosé-Hoover thermostat and a Parrinello-Rahman barostat with zero external pressure were applied throughout. The thermostatting and barostatting times were adjusted to be 0.1 and 1.0 ps, respectively, and the integration time step was chosen to be 1 fs. Figure 1(a) shows an example of the potential energy evolution of the total simulation cell during the hybrid Monte Carlo/molecular dynamics procedure. We count the system as relaxed when the corresponding energy gradient for the last 20 ps is less than 0.1 eV/ps (inset to Figure 1(a)). The equilibrated samples were analyzed and visualized using the OVITO software [43], in which dislocations were analyzed by the dislocation extraction algorithm [44] and the local atomic order around each atom was determined by the polyhedral template matching (PTM) method [45]. For atomic snapshots presented in this paper, solute atoms are shown as black while atoms of the matrix element are colored according to their local atomic order: face centered cubic (fcc) – green, body centered cubic (bcc) – blue, hexagonal close packed (hcp) – red, icosahedral – yellow and other – white. Atoms of the intermetallic phase $L1_2$ are shown in magenta.

Orthogonal simulation cells were created (Figure 1(b)) containing two edge dislocations of opposite character, with each full dislocation relaxing into two Shockley partial dislocations with a stacking fault in between. While the partial dislocations resulting from a full edge or full screw dislocation have different Burgers vectors, they are of mixed character in both cases, having both screw and edge components. This means that the local stress field, while quantitatively different (i.e., stress tensor components will have different magnitudes), will be qualitatively the same (i.e., have the same non-zero normal and shear stress components). The similarity between the stress tensors means that dislocations in fcc materials provide a consistent template for linear complexion formation. We performed additional simulations with dissociated screw dislocations and confirmed the same types of complexion transitions as for dissociated edge dislocations. Thus, for simplicity, we focus on reporting the results of simulations of the edge dislocation configuration in this manuscript. A consistent simulation cell was created for the various alloy systems by fixing the number of atomic planes in each direction. Each simulation cell is $60\sqrt{2}a$ in the X direction, $48\sqrt{3}a$ in the Y direction, and $10\sqrt{6}a$ in the Z direction, where $a$ is the face centered cubic lattice parameter corresponding to the base element of interest. Additional "long" simulation cells that extended 5 times further in the Z direction were used to verify the periodicity of the complexion structure along the line defect. The edge dislocations were inserted into the simulation cell by removing one half of the atomic plane in the middle of the sample and equilibrating the system with molecular statics. Creating two edge dislocations with opposite Burgers vectors enables periodic boundary conditions to be used in all directions.

## 3. Results and discussion

We explored local complexion transitions near dislocations in five different alloy systems, predicting three types of linear complexions: (1) nanoparticle arrays, (2) stacking fault complexions, and (3) platelet arrays. Thus, this section is divided into four subsections, three corresponding to a given complexion type and a fourth to discuss the possibility of multiple complexion coexistence. In the first subsection, we discuss in detail the formation of nanoparticles of an $L1_2$ intermetallic phase along dislocations in the Ni-Fe system. We then confirm the existence of similar $L1_2$ nanoparticle linear complexion arrays in long samples for Ni-Fe, Ni-Al, and Al-Zr, demonstrating that this type of behavior should be expected in many fcc alloys. In the second subsection, we discuss how solute segregation to stacking faults in Cu-Zr results in local chemical and structural transitions to the new two-dimensional, ordered features. We show that an increase in Zr composition leads to thickening of these features into three-dimensional objects and eventually a transformation to a disordered phase that completely destroys the original defect. This final step may be associated with a premelting transition or a bulk liquid phase depending on the temperature and local Zr composition. In the third section, we uncover the formation of two-dimensional platelets in the Al-Cu system that grow from the partial dislocations along the {100} crystalline planes.  These nanolayer features form in arrays along the dislocation lines and cause faceting of the dissociated dislocation, which is confirmed in long Al-Cu samples.  Finally, we demonstrate that more than one of these linear complexion types can be sustained in a ternary Al-Cu-Zr alloy.

*3.1.    Nanoparticle array linear complexions (Dislocation core unaffected)*

First, we explore linear complexion formation in fcc Ni-Fe alloys.  Figure 2(a) shows the pure Ni sample heated up to the target temperature of 500 K, in which extended edge dislocations have stacking fault widths of ~7-10 nm.  Some variation in stacking fault width is observed because the dislocations slightly interact with each other due to their stress fields.  Figure 2(b) shows the local hydrostatic stress for the same sample; only the partial dislocations cause local stress concentrations.  Generally, tensile stresses occur in the direction where the extra half-plane has been removed (i.e., in the middle of the sample) and compressive stresses occur in the other region (i.e., in the top and bottom sections of the cell).  Figure 2(c) presents a similar simulation cell but with 1 at.% Fe added, where the solute atoms segregate to the compression side of the partial dislocations and form nanometer-sized, ordered structures.  These linear complexions have the same crystallographic structure and orientation as the matrix phase and do not destroy the partial dislocations, which are still recognized by the DXA algorithm.  Additional analysis these particles with the PTM method indicates that the local atomic order and compositional patterning corresponds to the $L1_2$-$FeNi_3$ phase (see the embedded box in Figure 2(c)), which has coherent interfaces with the matrix phase in this case and grows into the viewing direction, along with the dislocation lines.  With progressive doping of Fe into the base Ni, two distinct particles of the $L1_2$-

FeNi$_3$ phase first form, one at each partial, then continue to grow as composition increases. Eventually, one large, coalesced particle is found above the partial dislocation-stacking fault collection. This process can be seen in Figure 2(c), where the bottom pair of partial dislocations has two distinct particles in the intermediate growth stage. The Fe atoms appear yellow in Figure 2(d), with the different appearance of the two particles resulting from the viewing plane cutting each at a different place, proving that they are distinct particles in the early stages. The top pair of partial dislocations shows the later stage of growth where two particles have coalesced to become one. Figure 2(e) shows an even higher dopant concentration where single particles are found at both sets of partial dislocation pairs. The particles are also larger on average than they were in the sample with less dopant. In all cases, the formation of the ordered structures relaxes the local hydrostatic compression stress, as shown in Figures 2(d) and (f). It is interesting to note that, with an increase in the Fe concentration and with the growth of the ordered complexions, the stacking fault width decreases. This promotes the merging of the nanoscale size particles formed at the opposite sides of the dissociated dislocation but also provides a testable hypothesis for future experiments. While local chemical ordering can be difficult to observe with transmission electron microscopy, a reduction in stacking fault width would be a sign that such a linear complexion transformation has occurred.

For a systematic investigation of linear complexion nucleation in Ni-Fe, we considered a wider range of compositions and temperatures. Equilibrium states with global compositions ranging from 1 to 8 at.% Fe and temperatures ranging from 300 to 800 K were created. The results are summarized in Figure 3(a), creating a section of the linear complexion diagram for Ni-Fe that is analogous to a bulk phase diagram. The samples which show no precipitation, where only a random solid solution fcc phase is observed, are denoted as black dots. Samples with nanoscale

L1$_2$-FeNi$_3$ precipitates are shown as magenta circles with the size of the data point proportional to the atomic fraction of the intermetallic phase in the simulation cell. Since the number of possible nucleation sites where precipitates can form in our computational cells is constant (the two dissociated dislocations), an increasing atomic fraction of the cell signals that particles are becoming larger. The average size of the precipitates increases with an increase in a global composition and/or a decrease in temperature. Depending on the annealing temperature and global composition, the local composition of the L1$_2$-FeNi$_3$ precipitates varies from 24 to 27 at.% Fe, values which are within the correct range for the L1$_2$ phase given by the bulk phase diagram for this potential [36]. The bounds of the two-phase (fcc + L1$_2$) coexistence zone on the bulk Fe-Ni phase diagram is shown in Figure 3(a) as a solid black line. The L1$_2$ particles that formed inside of this region are expected from the bulk phase diagram but just happened to form at the dislocation due to the local segregation of Fe atoms. This can be viewed as a case of dislocation-assisted heterogeneous nucleation. In contrast, the nanometer-sized precipitates outside of the two-phase coexistence region are stabilized by a dislocation' stress field and represent unique defect states associated with linear complexions. The border between the solid solution and linear complexion regions (dashed line in Figure 3(a)) can be defined by finding the solute saturation in the bulk of the lattice, away from the dislocations, which we fit with an Arrhenius formula $x = x_0 \cdot \exp(-\Delta G/RT)$ that accounts for the presence of dislocation stress field in the system [46, 47], where $x$ and $x_o$ are the bulk solubility limits in the presence and the absence of dislocations, respectively, $\Delta G$ is the free energy change due to the presence of dislocations, $R$ is the gas constant, and $T$ is the temperature. The bulk composition is simply the sample composition after removing the L1$_2$ complexions shown in Figure 3(b). Similar nanoparticle complexions were recently studied by Turlo and Rupert [26] on the opposite, Fe-rich side of the phase diagram, where nanometer-sized precipitate arrays comprised of a metastable B2-FeNi phase and a stable $L1_0$-FeNi phase were observed at the compression side of edge dislocation lines. Since the matrix material was bcc in that case, the dislocation did not dissociate and no particle coalescence was observed.

To explain why Fe atoms segregate to the compression side of the dislocation, even though they have a larger atomic radius and would traditionally be expected to prefer the tension side, we estimated an equilibrium atomic volume corresponding to the minimum potential energy for the three phases: fcc Ni, an fcc solid solution with composition Ni-25at.% Fe, and the ordered $L1_2$-FeNi$_3$ phase. The calculations were performed using the same simulation cell but without dislocations. The simulation box sizes were progressively rescaled to vary the average atomic volume in the range of 9 to 12.5 Å$^3$. Figure 3(c) shows the energy of the various phases as a function of average atomic volume, where it is obvious that the equilibrium atomic volumes of the solid solution alloy and of the intermetallic phase are smaller than that of the pure Ni. This preference for a smaller lattice spacing explains why the Fe solute atoms segregate to the compression side of dislocations. Another important finding from Figure 3(c) is that the equilibrium energy of the solid solution is much higher than that of the $L1_2$-FeNi$_3$. This difference in the bulk energies promotes the precipitation of the intermetallic phase, while the low-energy coherent interfaces with the matrix phase keep the corresponding nucleation barrier low for such a transition.

Additional analysis can also epxlain the reduction in stacking fault width with Fe segregation and linear complexion formation. Many theoretical and experimental investigations suggest stacking fault size can be described as $\delta \propto G/\gamma_{SF}$, where $\delta$ is the equilibrium stacking fault width, $G$ is the shear modulus, and $\gamma_{SF}$ is the stacking fault energy of the system [48]. The localized nature of the solute segregation and phase transformation near the dislocations means that such changes occur in an inhomogeneous manner around the stacking fault (see Figure 2). Several additional simulations were performed to measure how the alloying and different local inhomogeneities influence key parameters such as the shear modulus and the stacking fault energy, as shown in Figure 4(a). The simulation cell, identical to the one shown in Figure 1(a) but without dislocations, was strained in one set of simulations to obtain the shear modulus, while a stacking fault was inserted in another set to extract the stacking fault energy for each system. We considered four different systems: pure fcc Ni, a solid solution fcc phase with composition Ni-2 at.% Fe, a mixture of fcc Ni-4 at.% Fe and pure fcc Ni, and finally a mixture of $L1_2$-FeNi$_3$ and pure fcc Ni. As demonstrated in Figure 4(a), while both the shear modulus and the stacking fault energy increases with solute and intermetallic phase addition, their ratio decreases. This means that the equilibrium stacking fault width should decrease as well, which is consistent with our observations in Figure 2. Moreover, as shown in Figure 4(b), the homogeneous Ni-2 at.% Fe and inhomogeneous Ni-4 at.% Fe/fcc Ni systems, which have the same global composition, demonstrate the same values of the key parameters, meaning that the segregation and depletion around a dislocation core should not affect the equilibrium stacking fault width. Addition of Fe in either manner gives the same effect. However, the local complexion transformation to a nanoscale $L1_2$-FeNi$_3$ phase near the dislocations further reduces the equilibrium stacking fault width, making it a clear indicator of the complexion transition.

By equilibrating additional long samples, we confirm the existence of stable particle arrays that initial form at the partial dislocations in Ni-Al and Al-Zr, where Ni-rich and Al-rich $L1_2$ phases were found, respectively (see Figure 5(a-f)). The pure Ni sample for the Ni-Al potential (Figure 5(a)) has an extremely narrow stacking fault. This starting stacking fault is very different than what was observed for the pure Ni sample simulated with the Ni-Fe potential (Figure 2(a)), highlighting that all atomistic modeling observations are dependent on the interatomic potential used. In the case of Ni-Al and Al-Zr shown in Figure 5, the potentials were created with an emphasis on recreating the properties of the intermetallic phases, which may have resulted in some deviations for the pure metal properties. Figures 5(b) and (c) shows the final structures of Ni-2.5 at.% Al and Ni-4.0 at.% Al samples equilibrated at 400 K, where the stacking faults are slightly wider than in the pure Ni and nanoparticle arrays of the $L1_2$ phase are confined to the tension side of a pair of partial dislocations. The pure Al sample in Figure 5(d) has much wider stacking faults than any of the other systems discussed above, due to the much lower stacking fault energy in the simulated Al potential (again a deviation from reality, as experimental observations show that Al has one of the highest stacking fault energies among the fcc metals [49]). No matter the issues for the pure metal potentials, the fundamental observation of an interplay between the equilibrium stacking fault width and the linear complexion form persists. Multiple separate particles are found for the Al-Zr system (Figures 5(e) and (f)). Second phase precipitation and growth while maintaining coherent interfaces will cause a local stress increase due to a difference in the lattice parameters of the complexion and matrix phases. While this excess stress is partially relieved inside of the dislocation segregation zone, the stress/strain along the dislocation line would became too high if the whole segregation zone is filled by the second phase. Thus, the system introduces periodic segments of altering second phase and solid solution, as it is demonstrated in Figure 5(g). Taken as a whole, our findings from this section suggest the existence of a distinct type of linear complexion: nanometer-sized particle arrays constricted to the dislocation line, which we term ***nanoparticle array linear complexions***. This type of linear complexions preserves the original dislocations and is composed of stable (and/or metastable) phases that maintain low-energy interfaces with the matrix phase in all directions. The structure of this complexion appears to be simply determined by equilibrium thermodynamics, with the next phase on the bulk phase diagram being the chosen structure in most cases.

### 3.2. *Stacking fault linear complexions (Dislocation core delocalized)*

Next, we explored local chemical and structural transitions at dislocations in Cu-Zr. As shown in Figure 6(a), the pure Cu sample has an average stacking fault width of ~5 nm. Figure 6(b) shows the local atomic order and local stress distributions at the tension side of the dislocation slip plane indicated by the dashed purple box in Figure 6(a). As shown in Figure 6(b), the partial dislocations are separated far enough apart that their stress fields do not strongly overlap. The green lines in Figure 6(c) represent the hydrostatic stress profiles in the pure Cu sample in a 0.7 nm thick layer on the tension side of the two dislocations, showing that the average stress in the middle of the sample is small. Following our observations in Section 3.1 above, we began investigating relatively high temperatures and low Zr composition, at which only solute segregation is expected. Figure 6(d) presents the equilibrium distribution of Zr atoms (black) around the partial dislocation pairs for Cu-1 at.% Zr at 900 K, showing strong segregation on the tension side and slight depletion on the compression side of the dislocations. Similar to our Ni-Fe observations, doping reduces the stacking fault width (Figure 6(e)). In addition, solute segregation to the stacking faults is also observed (Figure 6(e), top panel). This effect can also be seen in the composition profiles (red lines) in Figure 6(f), where there is an extensive Zr segregation to the partial dislocations and a slightly less elevated Zr composition in the middle of the stacking fault. These trends in the composition are generally consistent with the corresponding hydrostatic stress profiles in Figure 6(c) (red lines), with higher tensile stresses leading to higher Zr concentrations.

These observations allow one to build a mechanistic description of the doping process. First, doping with Zr leads to the shrinkage of the stacking fault. This in turn leads to an overlap of the stress fields of the leading and trailing partial dislocations, attracting more solutes to the tension side of the stacking fault. A decrease in temperature to 800 K at the same composition leads to more intense solute segregation, as shown in Figure 6(g). The top panel in Figure 6(h) shows that this increased segregation causes parts of the stacking fault to no longer be indexed as hcp. Instead, local chemical ordering is found at the stacking fault and the stress fields associated with the original partial dislocation pair begin to be destroyed. At this point, the original partial dislocations were not recognized anymore by the DXA method. This ordered structure produces one concentration peak with a composition of ~20 at.% Zr, as shown by the blue curves in the bottom panel of Figure 6(f). Such an ordered two-dimensional structure cannot be associated with any bulk phase since it requires the stacking fault template. Here, we refer to these complexions as *stacking fault complexions*, since they keep the shape of the stacking fault. The new complexions have a planar geometry, similar to a grain boundary complexion, but were created due to segregation and transformation at line defects. Moreover, the new complexion is limited in its width while the length goes through the simulation cell. As such, we still refer to these features as linear complexions here, but acknowledge that this terminology should be discussed and evaluated further.

A further increase in the Zr composition leads first to the widening of the "ribbon-shaped" complexion, as shown in Figure 7(a). The local atomic structure of this complexion is shown in Figure 7(b) using the PTM method, where all of the Zr atoms are arranged in a hexagonal pattern, with each surrounded predominantly by Cu atoms with icosahedral atomic order. Figure 7(c) presents a radial distribution function (RDF) from the two-dimensional slice indicated by the yellow box in Figure 7(b), showing that the stacking fault complexion has a structure that is very similar to the structure of the (111) plane of the $Cu_5Zr$ intermetallic phase. While local strains associated with the nanoscale phase transformation smooth the peaks on the RDF profile corresponding to the ribbon complexion, several crystalline peaks can be still recognized. This phase is rich on icosahedral-ordered atoms (inset to Figure 7(c)) and the average local composition of the stacking fault complexions is ~16-19 at.% Zr, matching the nominal 16.7 at.% Zr expected from the $Cu_5Zr$ intermetallic phase. The experimental work of Peng et al. [50] on aging of a supersaturated Cu-Zr alloy also supports our findings. These authors showed that Zr-rich nanoscale clusters with disc-like shapes were formed on {111} planes and then $Cu_5Zr$ nanoscale precipitates are created by adding more {111} planes, forming coherent $\{111\}_{Cu_5Zr}\|\{111\}_{Cu}$ interfaces.

Thus, one might hypothesize that these stacking fault complexions with order similar to a slice from the $Cu_5Zr$ phase might act as a precursor to the formation of a bulk $Cu_5Zr$ phase or $Cu_9Zr_2$ phase. In fact, when one continues to increase the Zr concentration in the system, the stacking fault complexion thickens until it appears to look more like a wire than a ribbon. In Figures 8(a) and (b), the complexion has a noticeable thickness in both of the in-plane dimensions. However, the $Cu_5Zr$ intermetallic phase has a much higher atomic volume (13.6 Å$^3$) than fcc Cu (11.9 Å$^3$), meaning precipitation of bulk intermetallic precipitates would result in a significant strain energy penalty inside of the Cu lattice. Moreover, the different crystal structure and lattice constants of the $Cu_5Zr$ and $Cu_9Zr_2$ intermetallic phases would necessitate the formation of high-energy incoherent interfaces ($Cu_5Zr$ has a cubic lattice with $a = b = c = 0.6870$ nm while $Cu_9Zr_2$ has a tetragonal lattice with $a = b = 0.6856$ nm and $c = 0.6882$ nm [51]). Instead, continued doping with Zr or an increase in temperature leads to the formation of a disordered phase. Figure 8(c) shows an atomic snapshot from a specimen with 2.2 at.% Zr at 1000 K, where a second-phase precipitate is formed that destroys any sign of the original defects. The inset to Figure 8(c) shows a radial distribution function for the atoms inside this phase, demonstrating the lack of long-range order consistent with an amorphous structure. The nanoscale transition to a disordered phase might be energetically favorable due to the relaxation of the strain energy of destroyed dislocations. However, it is possible that complete relaxation of the system and release of any strain energy penalty would lead to the formation of an intermetallic particle. However, we do not explore this possibility in any more details, since the behavior near the dislocations is our focus in this study, not transformation pathways to bulk phases.

To classify and analyze the variety of structures obtained in the Cu-Zr system, we construct a linear complexion phase diagram in Figure 9(a). Three regions were identified corresponding to samples with (1) solute segregation but no phase transformation (black dots), (2) ordered complexion transformations (yellow circles) into stacking fault complexions, and (3) heterogeneous precipitation of a disordered phase (purple circles). The solid line in Figures 9(a) represents the bulk melting curve, obtained for the interatomic potential used in this work [52]. Figure 9(b) shows the data replotted against the local composition, rather than the global composition, showing that the samples at 1200 K are locally above the melting curve and thus correspond to a bulk liquid phase. However, other samples with disordered precipitates are below the bulk melting curve and thus correspond to a premelting transition that does not appear on the bulk phase diagram. Hence, these structures are complexions, as they would not exist if it were not for the dislocations. It is interesting to note that the local compositions of the disordered complexions do not overlap with the compositions of the ordered complexions, indicating an existence of a phase boundary for the complexion transition, which is schematically shown as a dashed line in Figure (b). However, such a transition is not expected to be reversible because the formation of ordered complexions requires the presence of the original defect.

Finally, we confirm the existence of stacking fault linear complexions using a long simulation cell for Cu-1 at.% Zr equilibrated at 800 K. Figure 10 shows the perspective, top, and zoomed views of this system, where the stacking fault complexion transformation happens all along the dislocation line. As shown in Figure 10(c), the final structure of the linear complexion is made of the wide chemically ordered zones comprised of Zr atoms surrounded by Cu atoms with icosahedral atomic order, connected by narrow "necking" zones of different local atomic ordering. Such necking zones might represent the intermediate transition state from the original hcp structure of the stacking fault to the icosahedral structure of the stacking fault complexion. The stacking fault complexions represent an extreme type of linear complexion transition, where the original dislocation cores are delocalized, even though they were instrumental in driving the transition.

### 3.3 *Platelet array linear complexions (Dislocation core restructured)*

Figure 11(a) shows an equilibrated simulation cell for Al-0.1 at.% Cu at 300 K. The relatively large spacing between partial dislocations means that their stress fields do not overlap, similar to what was observed in Cu-Zr before doping. Due to the low atomic volume of Cu relative to Al, the Cu atoms segregate to the compression side of the partial dislocations while no noticeable segregation to the stacking faults was observed in this alloy system. Lowering the simulation temperature or increasing dopant composition promotes more intense segregation in the form of local clustering of Cu atoms along the dislocation lines (Figure 11(b)). The Cu atoms cluster in the shape of half of a disk, occupying only one atomic plane while causing a local structural transition that appears as bcc structural ordering according to the PTM method (see Figure 11(c)). This structural transition shifts the partial dislocation lines out of their glide planes in the growth direction of the bcc planes, with the dislocations identified by DXA shown as green lines in Figure 11(c). In addition, this structural transition causes a faceting of the partial dislocation lines that can be sen in Figures 11(b) and (e).

Made of one Cu plane, these features resemble as the well-known Guinier-Preston (GP) zones. GP zones are usually observed during the ageing of Al-Cu binary alloys and appear as thin platelets with a {100} orientation [53], as a precursor to precipitation. A closer look at the growth plane (Figures 11(d)) demonstrates that, like GP zones, the clusters form coherent interfaces with the matrix phase along the (100) plane. GP zones can also be identified by a significant difference in the local atomic volume, a feature again shared with the platelets in Figure 11(d). Figure 11(e) presents the equilibrium structure of a long simulation cell of Al-0.3 at.% Cu equilibrated at 300 K, which confirms the formation of arrays of these nanoscale-size layered features. Based on these observations, we denotes these linear complexions as *platelet array linear complexions*. The result of this complexion transition is a regularly faceted dislocation line. As these GP zones are made of just one Cu layer, they are usually referred to in the literature as GPI zones or a $\theta''$ phase [54]. Our atomistic simulations are in agreement with observations by Liu et al. [55] of aging treatments in a supersaturated Al-Cu alloy, in which the formation of arrays of GP zones was found along the dislocation lines.

A further increase in a global composition leads to the expansion of these platelet structures, but growing slightly in thickness but also extending further into the material, as demonstrated in Figure 12(a). Figure 12(b) shows the slice perpendicular to the platelet complexions, demonstrating that the maximum thickness is three atomic planes, comprised of two

Cu planes with one Al plane in between. These layered features are thicker than GPI zones but still cannot be associated with the bulk phase due to their two-dimensional nature. Similar nanolayer platelets in the Al-Cu system are usually referred to in the literature as GPII zones or a $\theta'$ phase, while their precise structure is still debated [54]. The platelet complexions observed in this figure are comprised of the three (001) atomic planes of the $\theta - Al_2Cu$ phase (*Fm3m* space group, see Figure 13(a)). The magenta line in Figure 13(a) represents a coordination analysis of the trilayer feature shown in Figure 12, matching well the RDF for the $\theta - Al_2Cu$ phase (black line in Figure 13(a)). These trilayer features cause faceting of the stacking fault in the same way as one layer features observed at lower compositions (Figure 11), and would also be expected to form arrays along the dislocation line. Such arrays of nanoscale trilayer features can be also denoted as platelet array linear complexions. However, if these features expand into the bulk far beyond the dislocation stress field, they can exist independently from the defect and should not be considered as a part of a complexion. Rather, the original linear complexion has served as a site for heterogeneous nucleation of a new phase. A range of temperatures and compositions were modeled to construct an Al-Cu linear complexion diagram that is presented in Figure 13(b). There are no bulk phase transition lines on this figure and the dotted line schematically defines the border for the complexion transition. Figure 13(b) shows that the atomic fraction (and therefore the complexion size) increases with an increase in global composition or a decrease in temperature. The local complexion composition mainly stays in the range of 60-70 at.% Cu, with some small complexions at low temperatures and low compositions reaching local compositions as high as ~88 at.% Cu.

### 3.4 *Coexistence of multiple complexion types*

As the number of dopant elements in an alloy increases, it is possible that multiple types of linear complexions could be formed. Consider, for example, the case of dislocation in Al sample doped with both Zr and Cu. Zr atoms segregate to the tension side of partial dislocations, making linear complexions in a form of nanoscale particle arrays of the $L1_2$-$Al_3Zr$ phase (Figure 5(e)). On the other hand, Cu atoms segregate to the compression side of partial dislocations, making linear complexions in the form of platelet arrays of Cu-rich GP zones (Figure 11(e)). As these two complexion states are formed in different regions, it is reasonable to hypothesize that they can coexist in a ternary Al-Cu-Zr alloy. To test this hypothesis, additional Monte Carlo/molecular dynamics simulations were performed on the ternary Al-Cu-Zr system, with the same potential used for our studies of the binary constituents [39].

As representative complexion states were previously observed in Al-0.3 at.% Cu and Al-4.5 at.% Zr at 300 K, a first logical attempt would be to investigate the ternary Al-0.3 at.% Cu-4.5 at.% Zr system equilibrated at the same temperature. Figure 14 shows the equilibrium atomic configuration for this sample, where again linear complexions that are confined at dislocations are observed. Similarly to the binary Al-Cu system, GP zones are formed on the dislocation compression side along the partial dislocation lines, resulting in faceting of the stacking fault. However, the Zr-rich $L1_2$ phase precipitates do not follow the partial dislocation line as observed in the binary Al-Zr system, but instead prefer to grow near the GP zones. While both complexion form as expected, the linear complexions appear to be dependent on each other. Thus, linear complexions can have higher level complexity in a manner which is similar to interfacial complexions, when interactions of multiple segregating species to crystalline defects can generate new phenomena [56].

## 4. Conclusions

In this work, we performed atomistic simulations to explore linear complexions at dislocations in Ni-Fe, Ni-Al, Al-Zr, Cu-Zr, and Al-Cu fcc alloys. The corresponding linear complexion diagrams were constructed by investigating wide ranges of temperatures and compositions for each system. The linear complexions as well as the heterogeneously-nucleated bulk phases explored in this work are summarized in Figure 15, which also schematically define the place of linear complexions among other phases on a bulk phase diagram. The existence of three different types of linear complexions is predicted in this work: (1) nanoparticle array linear complexions in Ni-Fe, Ni-Al, and Al-Zr, (2) stacking fault linear complexions in Cu-Zr, and (3) platelet array linear complexions made of GP zones in Al-Cu. These three complexion types can be differentiated by their effect on the original dislocation core. For example, the nanoparticle array linear complexions leave the dislocation core unaffected, while the platelet array linear complexions restructure the dislocation core, forming facets comprised of edge and screw dislocations. Finally, the stacking fault linear complexions delocalize the dislocation core into a more relaxed structure, so it is no longer localized in space. Multiple linear complexion coexistence was also predicted in the ternary Al-Cu-Zr, with nanoscale arrays of GP zones coincident with $L1_2$-$Al_3Zr$ nanoparticles. In addition, a nanoscale complexion transition that completely destroys the original defect was observed in the Cu-Zr system, eventually leading to an amorphous structure. This disordered phase was formed well below the melting temperature and represents a nanoscale premelting transition.

By investigating several model systems, some fundamental conclusions regarding linear complexions and their types can be derived. The complexion region on a phase diagram is determined by the solubility limit curves in the presence and absence of dislocations. Nanoparticle array linear complexions are composed of phases that are able to form low-energy interfaces (sometimes these are coherent interfaces) with the matrix phase in all directions. If coherent interfaces can be maintained only at specific atomic planes, the layered complexions such as stacking fault linear complexions and platelet array linear complexions can be formed. Stacking fault linear complexions are expected to form if low-energy interfaces are maintained parallel to the dislocation slip planes ({111} for face-centered cubic metals), while platelet array linear complexions are expected to form in other cases. It is important to note that the complexion structure is typically the same as the next phase on the bulk phase diagram, with some metastable phases also appearing due to the limited size and/or shape near the dislocation. Taken as a whole, our work demonstrates that linear complexions are an intriguing new class of defect states, which allow for the creation of new materials comprised of networks of nanoscale-size complexions that are in thermodynamic equilibrium with the matrix phase.

**Acknowledgments:**

This research was supported by the U.S. Army Research Office under Grant W911NF-16-1-0369.

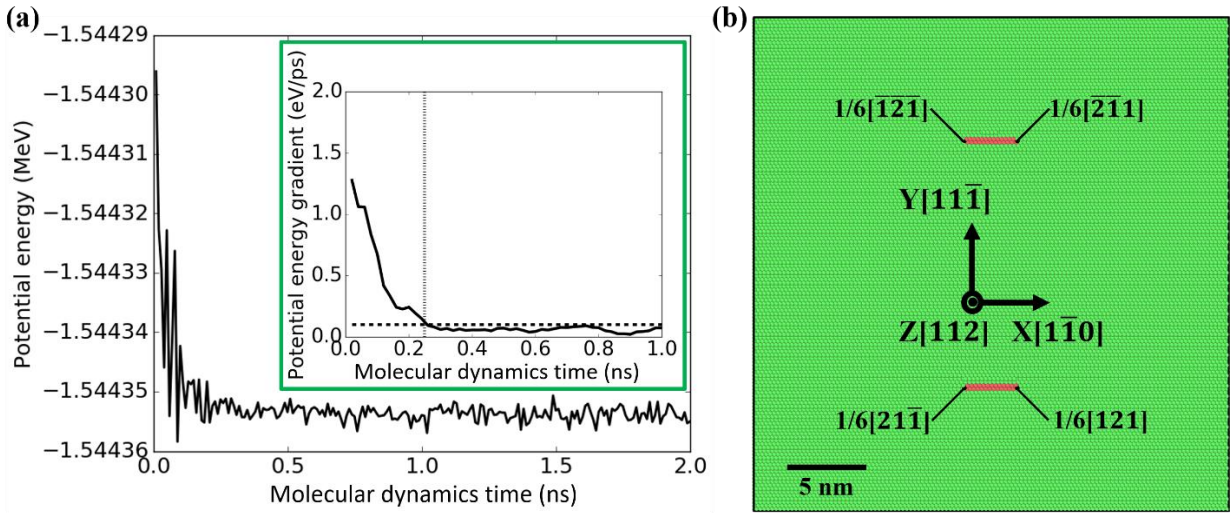

Figure 1. (a) The time evolution of the potential energy during the equilibration of a Ni-2 at.% Fe sample at 400 K. The embedded green box shows the potential energy gradient, with the dotted line denoting the threshold value of 0.1 eV/ps. (b) XY view of the initial system simulation cell with edge dislocations. Atoms are colored according to their local atomic order: green – fcc and red – hcp.

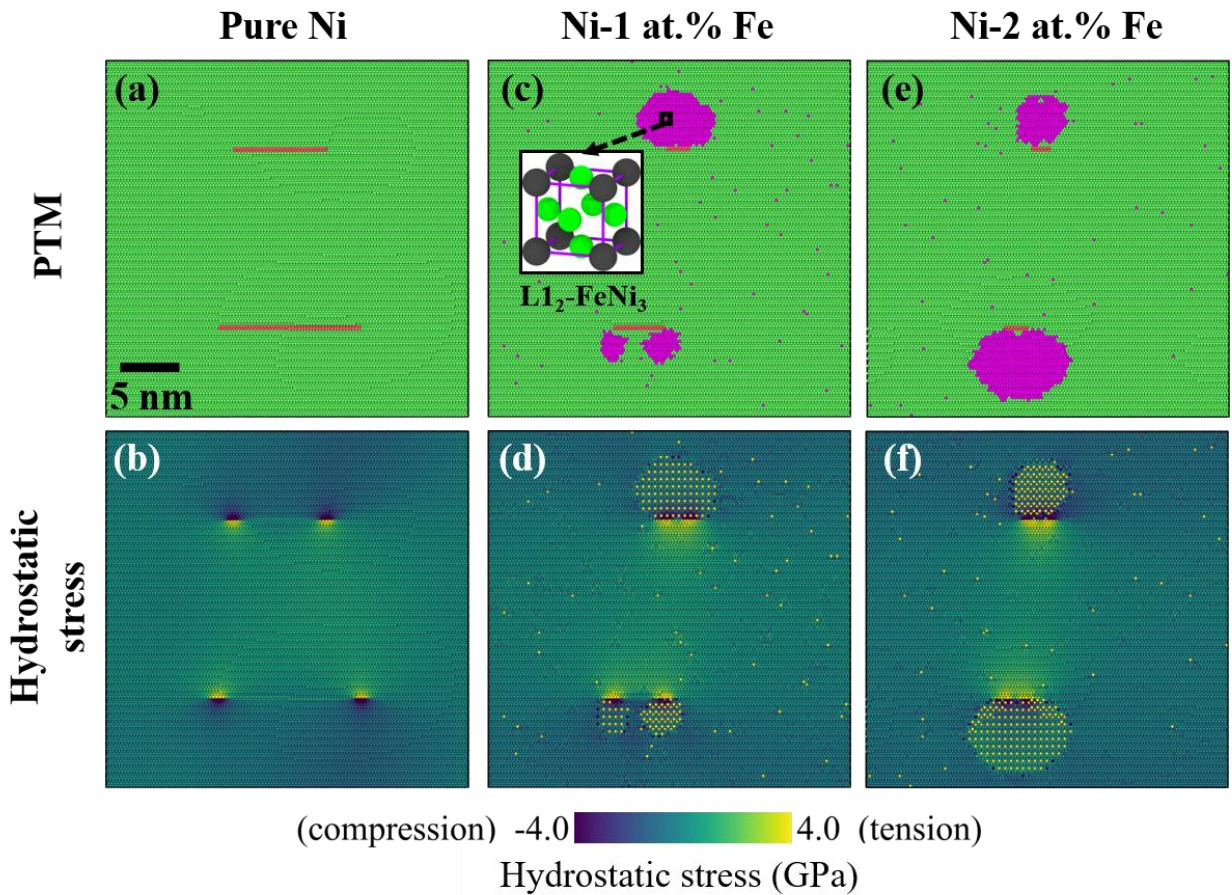

Figure 2. XY atomic snapshot of the (a,b) pure Ni, (c,d) Ni-1 at.% Fe, and (e,f) Ni-2 at.% Fe samples at 500 K. In (a,c,e), atoms are colored according to their local crystal structure obtained by the PTM method, with the atoms corresponding to the $L1_2$ phase shown in magenta. The embedded box in (c) demonstrates the chemical order of the $L1_2$ phase. In (b,d,f), atoms are colored according to the local hydrostatic stress.

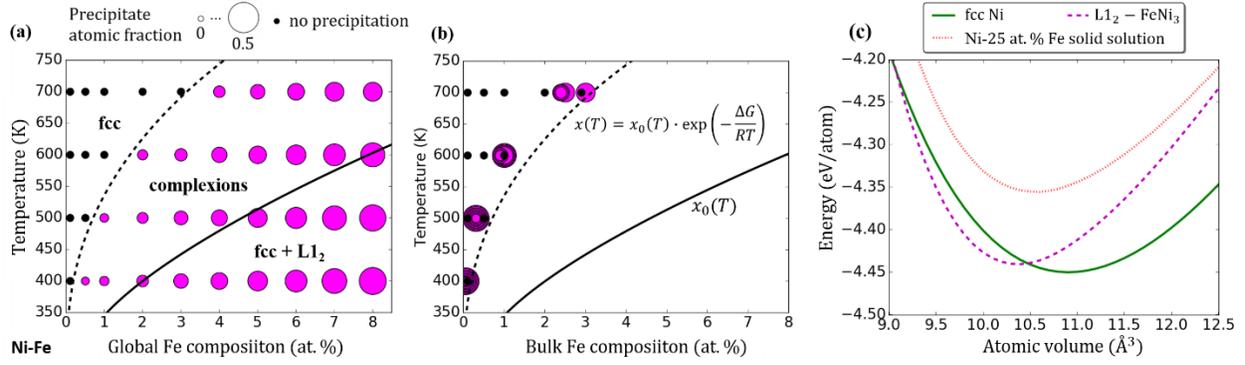

Figure 3. (a) Linear complexion diagram representing the conditions at which formation of the nanoscale intermetallic phase is observed (magenta circles) and not observed (black circles). The radius of the magenta circles corresponds to the atomic fraction of the $L1_2$-$FeNi_3$ phase in the simulation cell. The solid line shows the limit of the two-phase coexistence region on the bulk Ni-Fe phase diagram, computed for the potential used in this work [36]. The dashed line represents the solubility limit, $x$, in a presence of dislocations, obtained by adjusting the original solubility limit $x_o$ (solid line) using an Arrhenius formula: $x = x_0 \cdot \exp(-\Delta G/RT)$. (b) The solubility limit in the presence of dislocations is obtained by calculating the saturation of the bulk region. (c) Energy per atom as a function of the atomic volume measured at 0 K for three different phases listed on top of the figure.

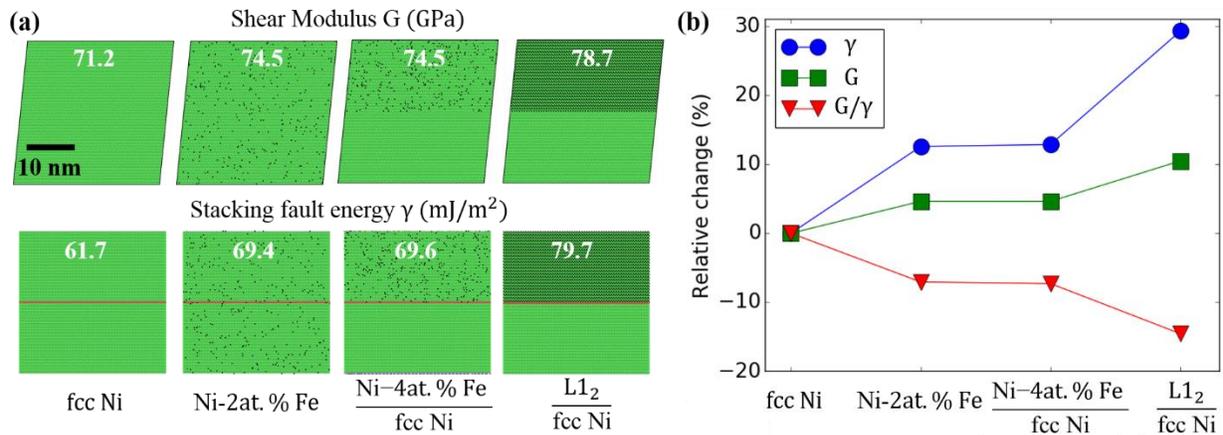

Figure 4. (a) XY snapshots of the simulation cells used to determine shear modulus (top) and stacking fault energy (bottom) of the four different configurations. Ni atoms are colored according to their local crystal structure obtained by the PTM method. Fe atoms are shown in black. (b) The relative change of the shear modulus ($G$), the stacking fault energy ($\gamma$), and their ratio using pure fcc Ni as the reference state.

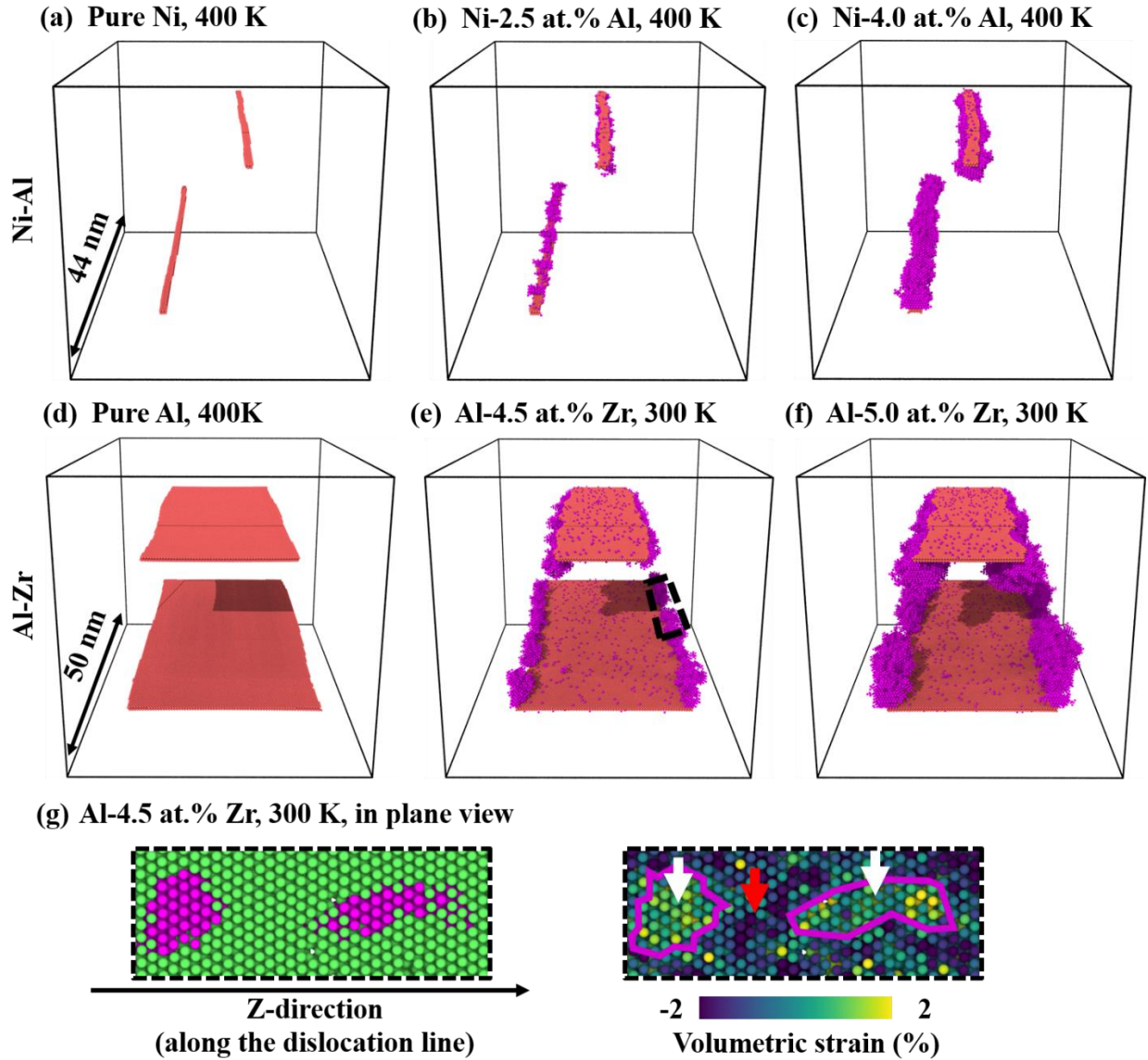

Figure 5. Perspective views of long simulation cells demonstrating nanoparticle arrays along the dislocation lines in (a-c) Ni-Al, and (d-f) Al-Zr. The atoms corresponding to an fcc solid solution were removed to increase the visibility of linear complexions. Red color shows stacking faults (hcp structure) and magenta color denotes an $L1_2$ phase. (g) A slice cut parallel and 2 nm above the slip plane of the area highlited as a dashed box in (e) with atoms colored according to their chemical ordering: green – fcc solid solution, magenta - $L1_2$ phase (left), and volumetric strain (right). Magenta lines highlight the limits of $L1_2$ precipitates, while white and red arrows indicate the regions of positive and negative strain, respectively.

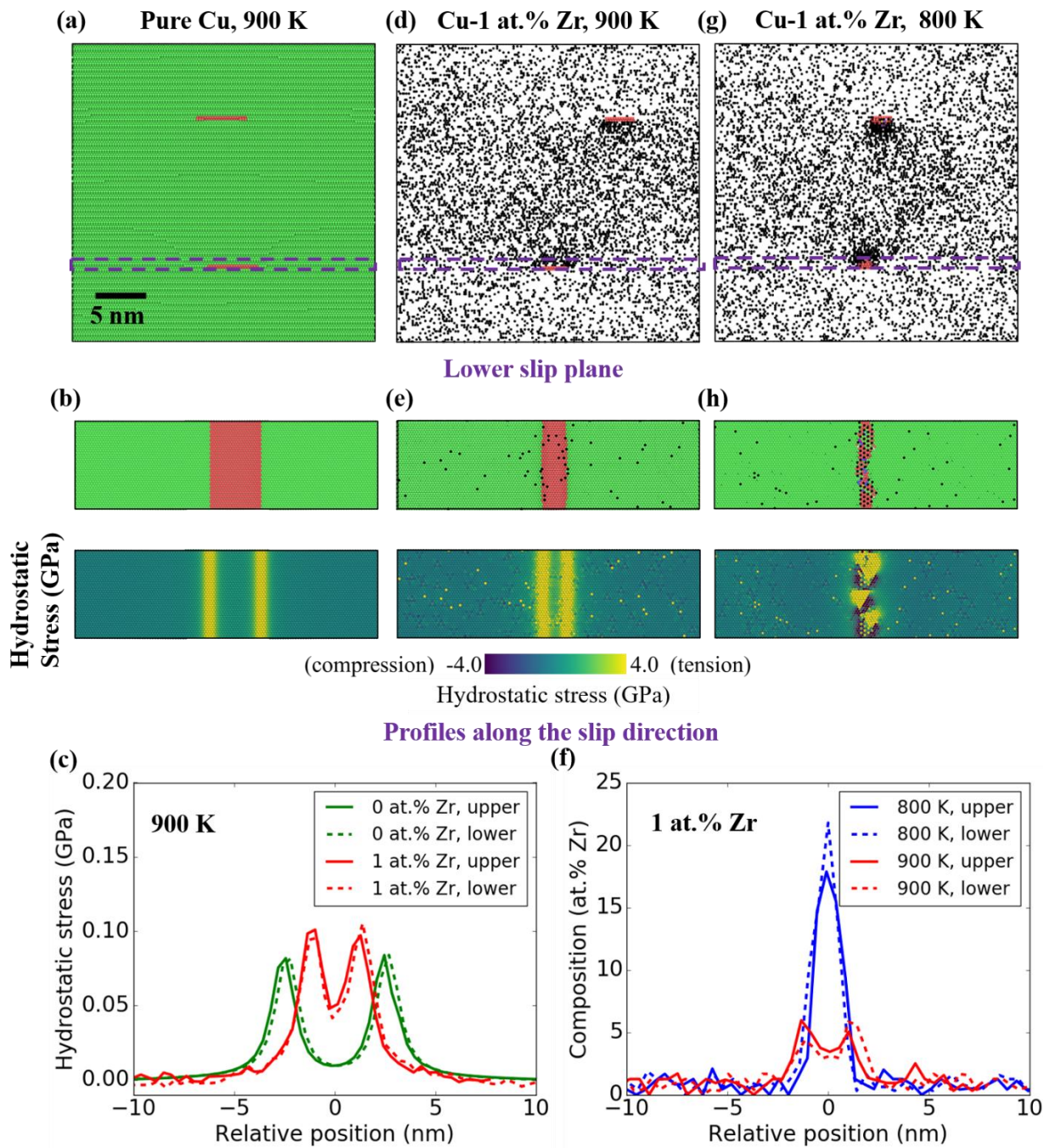

Figure 6. The cross-section and the bottom slip-plane views of (a,b) pure Cu at 900 K, (d,e) Cu-1 at.% Zr at 900 K, and (g,h) Cu-1 at.% Zr at 800 K. Cu atoms are colored according to their local crystal structure obtained by the PTM method, while Zr atoms are shown in black. The fcc Cu atoms are removed from (d) and (g) to demonstrate the Zr redistribution around dislocations. Atoms in the bottom frames of (b,e,h) are colored according to the local hydrostatic stress. (c) The hydrostatic stress profiles and (f) local composition profiles along the dislocation slip direction in the 0.7 nm layer next to the slip plane on the tension side of the dislocation core (shown as a violet dashed box in (a), (d), and (g)).

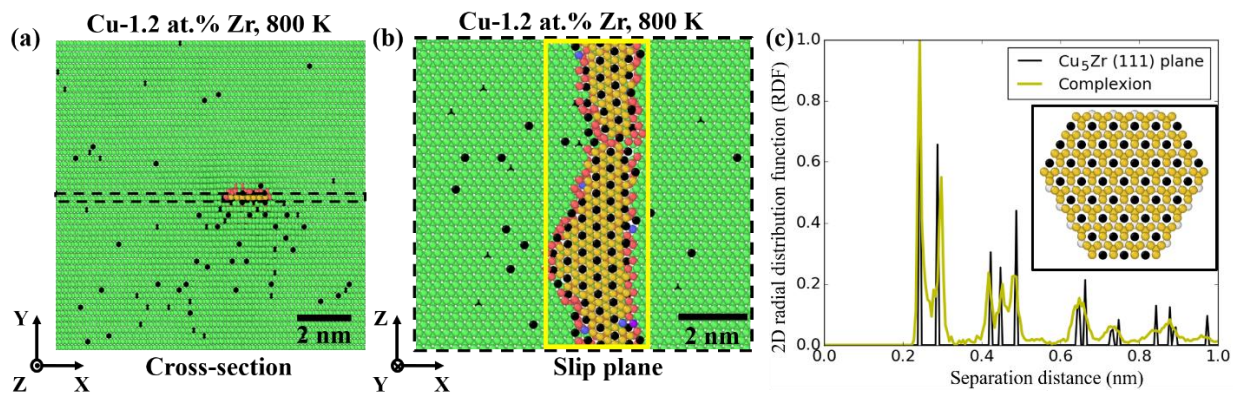

Figure 7. (a) Cross-sectional and (b) slip plane views of the stacking fault complexion in Cu-1.2 at.% Zr at 800 K. (c) Normalized radial distribution functions of the stacking fault linear complexion, selected as shown by a yellow box in (b), and a (111) plane of the $Cu_5Zr$ phase (see the embedded box in (c)). Cu atoms are colored according to their local atomic structure (blue – bcc, green – fcc, red – hcp, yellow – icosahedral, and light grey - other), while Zr atoms are colored black.

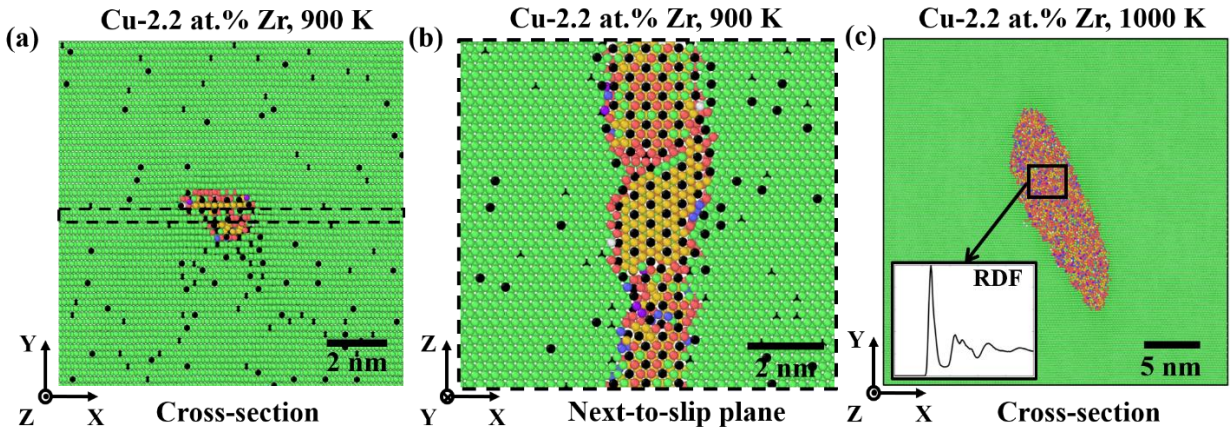

Figure 8. (a) Cross-sectional and (b) in-plane views of the stacking fault complexion in Cu-2.2 at.% Zr at 900 K. (c) Cross-sectional view of the disordered complexions in Cu-2.2 at.% Zr at 1000 K. The Cu atoms are colored according to their local atomic structure (blue – bcc, green – fcc, red – hcp, yellow – icosahedral, and light grey - other), while Zr atoms are colored black. The inset in (c) demonstrates the radial distribution function for the disordered phase.

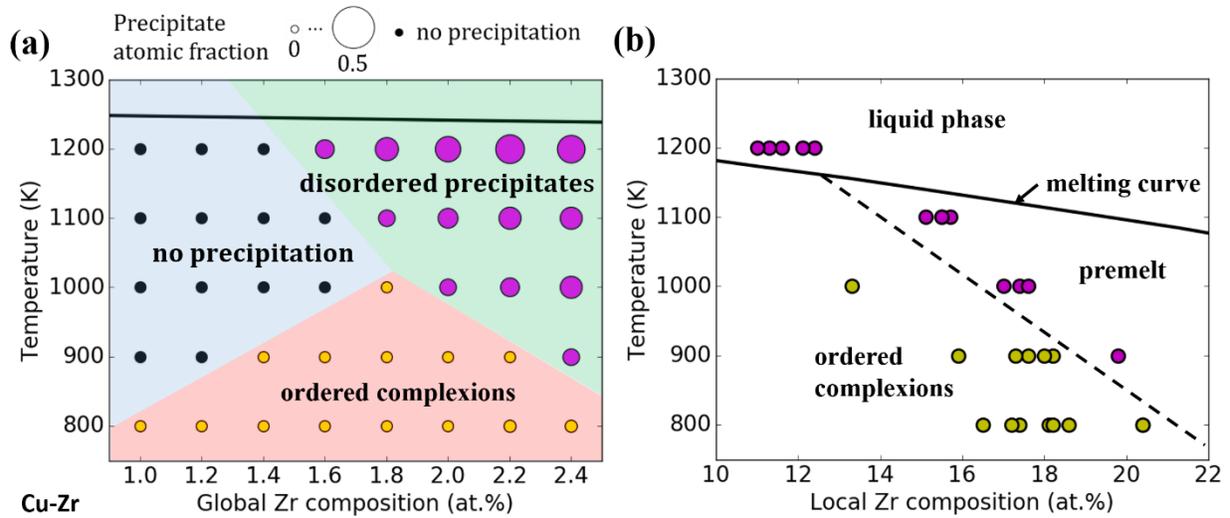

Figure 9. Linear complexion diagrams for the Cu-Zr system represented in terms of (a) global and (b) local (complexion) compositions. Ordered complexions are shown in yellow, while disordered complexions and bulk phases are shown in magenta. The solid black line represents the melting curve for Cu-Zr computed for the interatomic potential used in this work [52]. The dashed line was added to schematically demonstrate that the local compositions of ordered and disordered complexions can be separated from each other and do not overlap.

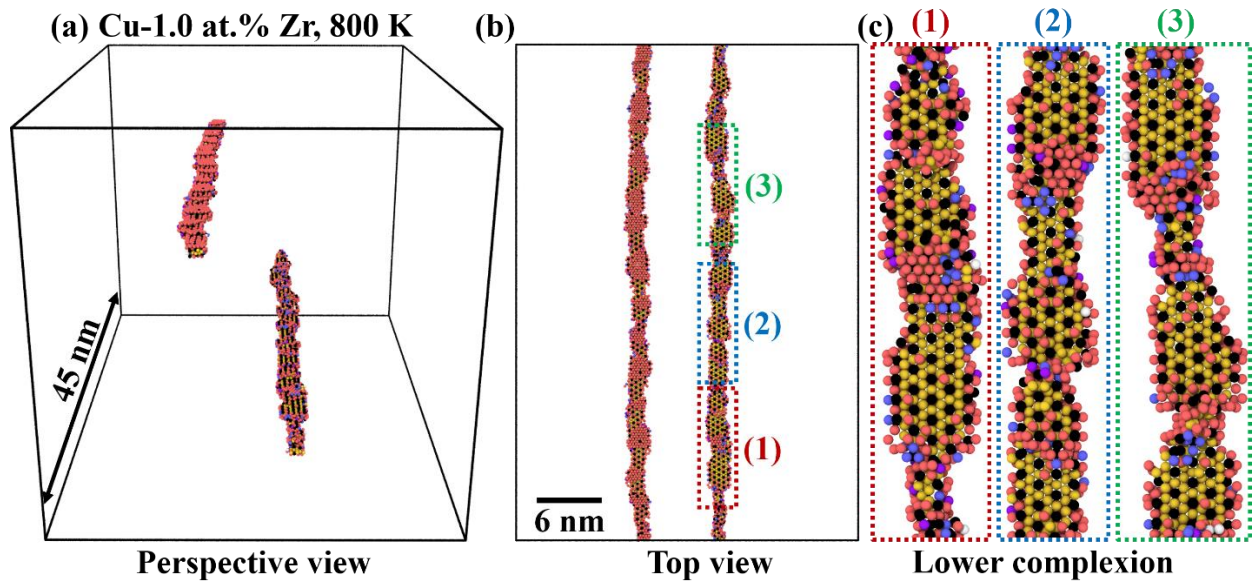

Figure 10. (a) Perspective and (b) top views of the long Cu-1 at.% Zr sample equilibrated at 800 K. (c) Zoomed regions of the lower dislocation indicated by dashed boxes in (b). The atoms corresponding to an fcc solid solution were removed to increase the visibility of linear complexions. Cu atoms are colored according to their local atomic order and Zr atoms are shown in black.

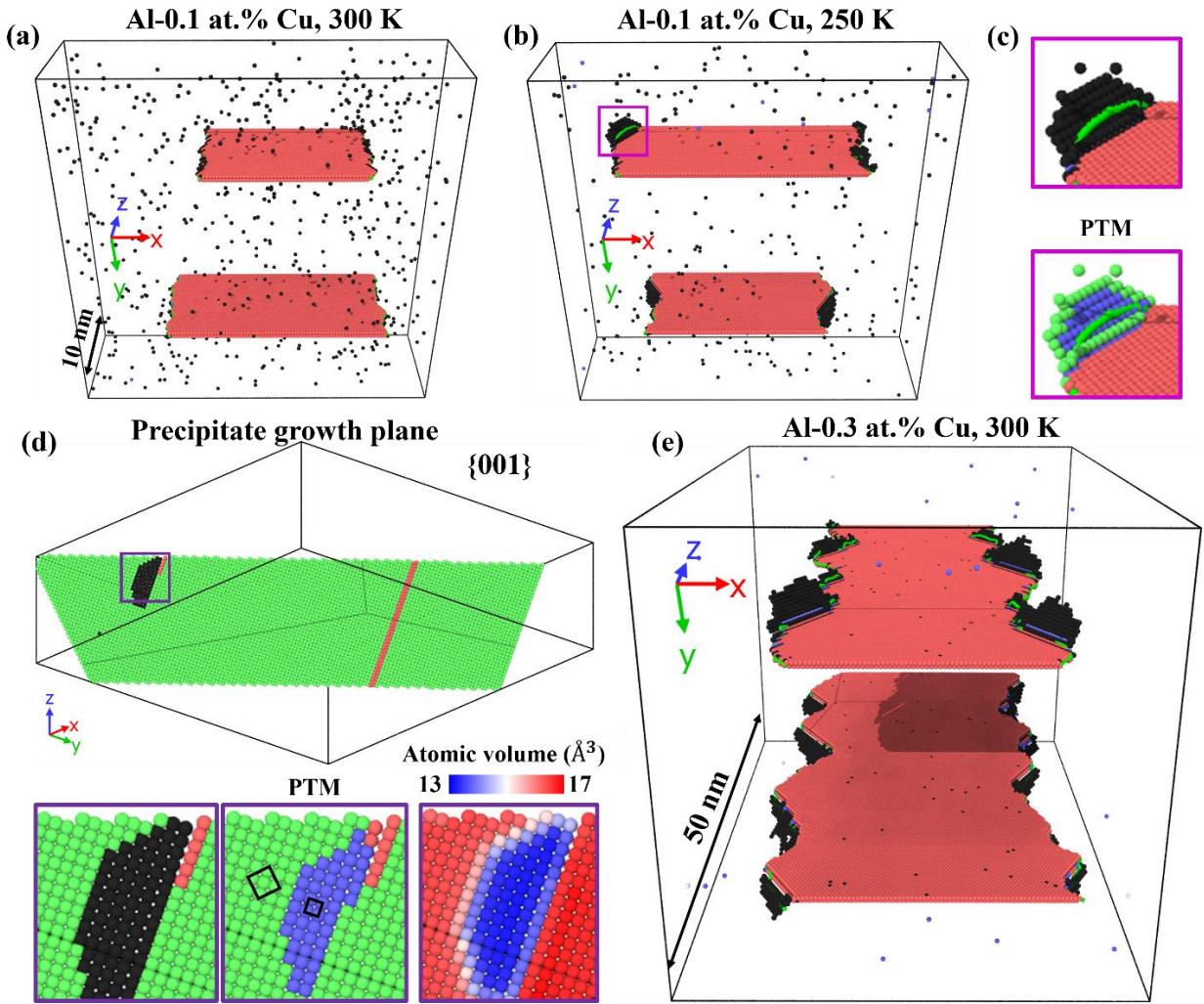

Figure 11. Perspective view of (a) Al-0.1 at.% Cu at 300 K, (b-d) Al-0.1 at.% Cu at 250 K, and (e) a long simulation cell of Al-0.3 at.% Cu at 300 K. The Al atoms are colored according to their local atomic structure, while Cu dopant atoms are shown in black. The fcc Al atoms are removed from the snapshots (a-c) and (e). Partial dislocations, identified by the DXA method, are shown as green lines.

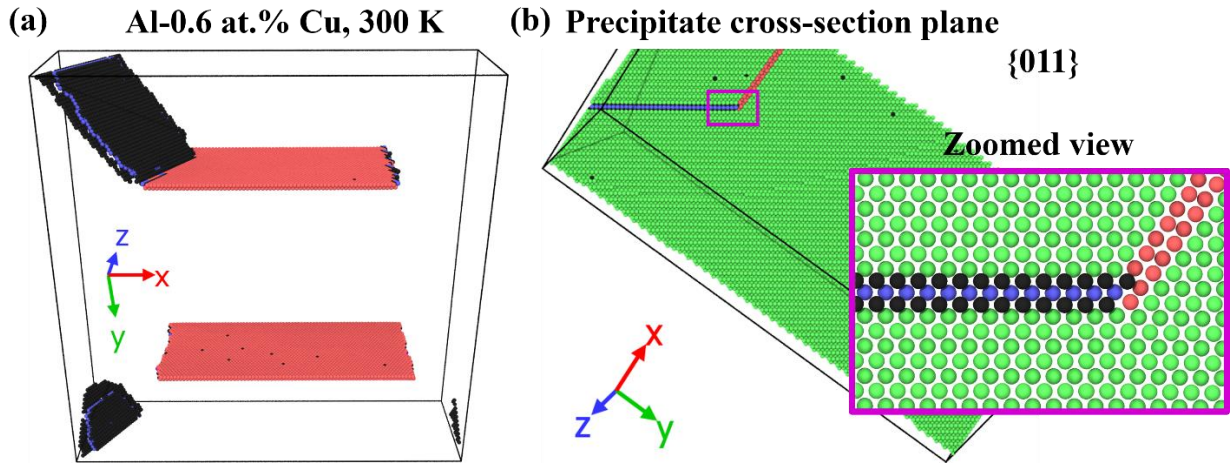

Figure 12. (a) Perspective view of Al-0.6 at.% Cu equilibrated at 300 K. (b) The precipitate cross-section plane with the zoomed view shown by the magenta box. Al atoms are colored according to their local atomic structure, while Cu atoms are shown in black.

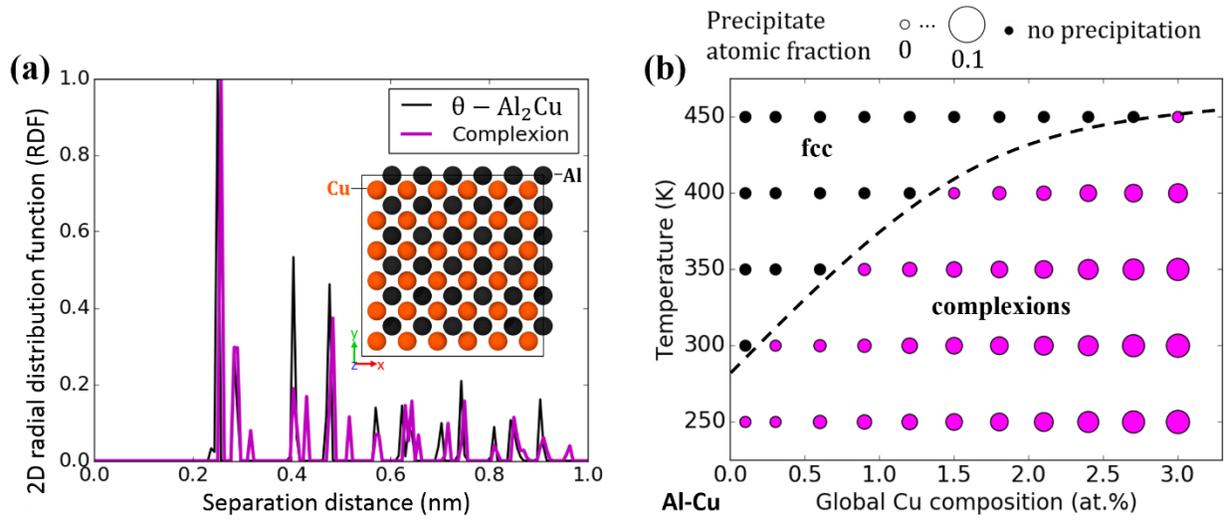

Figure 13. (a) Normalized radial distribution functions of the nanolayer linear complexion shown in Figure 12(a) and the $\theta - Al_2Cu$ phase (inset shows the atomic structure). (b) Linear complexion diagram for the Al-Cu system, in terms of global composition and temperature. The dashed line schematically indicates the boundary for the complexion transition.

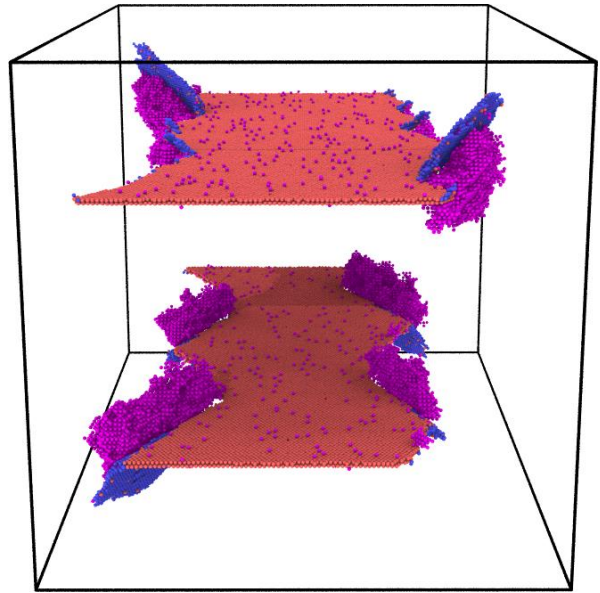

Figure 14. Perspective view of the long Al-0.3 at.% Cu-4.5 at.% Zr sample equilibrated at 300 K. The fcc Al atoms are removed from the snapshot. Red color corresponds to stacking faults (hcp atoms), magenta color corresponds to an $L1_2$ phase, and blue color corresponds to GP zones.

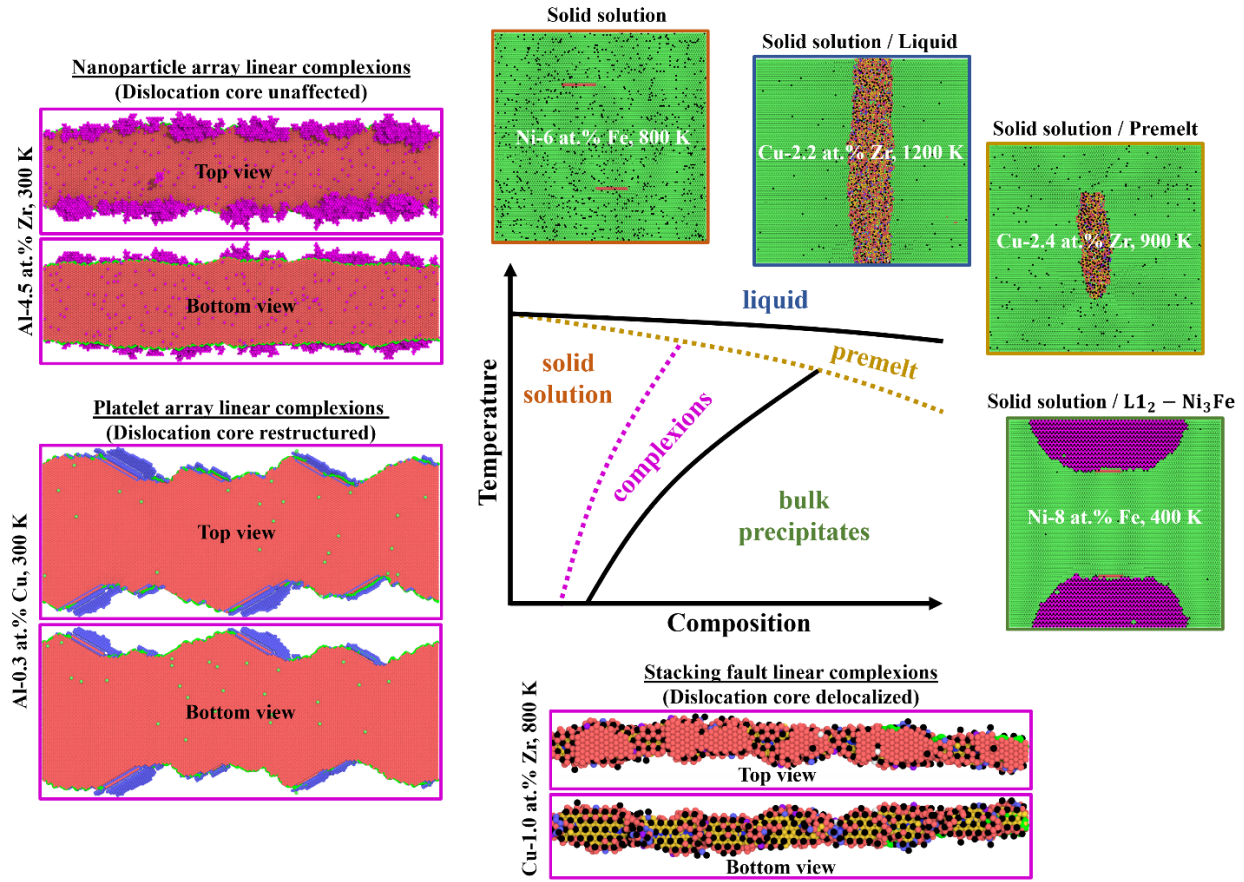

Figure 15. Schematic representation of all of the linear complexions discussed in this work, as well as the bulk phases that were probed. The middle frame defines the location of linear complexions among other features on a bulk phase diagram.